\documentclass[fleqn,usenatbib]{mnras}

\usepackage{newtxtext,newtxmath}
\usepackage{threeparttablex}

\usepackage[T1]{fontenc}

\DeclareRobustCommand{\VAN}[3]{#2}
\let\VANthebibliography\thebibliography
\def\thebibliography{\DeclareRobustCommand{\VAN}[3]{##3}\VANthebibliography}

\usepackage{graphicx}	
\usepackage{amsmath}
\usepackage{subfigure} 

\usepackage[normalem]{ulem}  
\newcommand\redout{\bgroup\markoverwith
{\textcolor{red}{\rule[0.5ex]{2pt}{0.8pt}}}\ULon}

\usepackage{tikz,xcolor,hyperref}

\title[Simulated non-thermal emission from SNR G1.9+0.3]{Simulated non-thermal emission of the supernova remnant G1.9+0.3}

\author[Villagran, M. A. et al.]
{M. A. Villagran$^{1}$\thanks{mvillagran@iafe.uba.ar},
D. O. Gómez$^{1}$, 
P. F. Velázquez$^{2}$,
D. M.-A. Meyer$^{3}$,
A. Chiotellis$^{4}$,
A. C. Raga$^{2}$,
\and
A. Esquivel$^{2}$,
J. C. Toledo-Roy$^{2}$,
K. M. Vargas-Rojas$^{2}$,
and E. M. Schneiter$^{5}$
\\
$^{1}$ Instituto de Astronom\'{i}a y F\'{i}sica del Espacio (IAFE), Av. Int. G\"uiraldes 2620, 
               Pabellón IAFE, Ciudad Universitaria, 1428, Buenos Aires, Argentina \\
$^{2}$ Instituto de Ciencias Nucleares, Universidad Nacional Aut\'onoma de M\'exico, Ap. 70-543, CDMX, 04510, M\'exico\\
$^{3}$ Universit\" at Potsdam, Institut f\" ur Physik und Astronomie, 
                 Karl-Liebknecht-Strasse 24/25, 14476 Potsdam, Germany\\
$^{4}$ Institute for Astronomy, Astrophysics, Space Applications and Remote Sensing, 
                 National Observatory of Athens, 15236, Penteli, Greece \\
$^{5}$ Departamento de Materiales y Tecnolog\'{i}a, FCEFyN-UNC, Av. V\'elez Sarsfield 1611, C\'ordoba, Argentina 
}

\date{Accepted XXX. Received YYY; in original form ZZZ}

\pubyear{2015}

\begin{document}
\label{firstpage}
\pagerange{\pageref{firstpage}--\pageref{lastpage}}
\maketitle

\begin{abstract}
Supernova remnants are the nebular leftover of defunct stellar environments, resulting 
from the interaction between a supernova blastwave and the circumstellar medium shaped by 
the progenitor throughout its life. They display a large variety of non-spherical morphologies such as ears that shine non-thermally.
We have modelled the structure and the non-thermal emission of the supernova remnant G1.9+0.3 through 3D magnetohydrodynamic numerical simulations. We propose that the peculiar ear-shaped morphology of this supernova remnant results from the interaction of its blast wave with a magnetized circumstellar medium, 
which was previously asymmetrically shaped by the past stellar wind emanating from the progenitor star or its stellar companion. We created synthetic non-thermal radio and x-ray maps from our simulated remnant structure, which are in qualitative agreement with observations, forming ears on the polar directions. Our synthetic map study explains the discrepancies between the measured non-thermal radio and X-ray surface brightness distributions assuming that the Inverse Compton process produces the observed X-ray emission. 
%
\end{abstract}



\section{Introduction}

\begin{figure}
    \centering
    \includegraphics[width=\columnwidth]{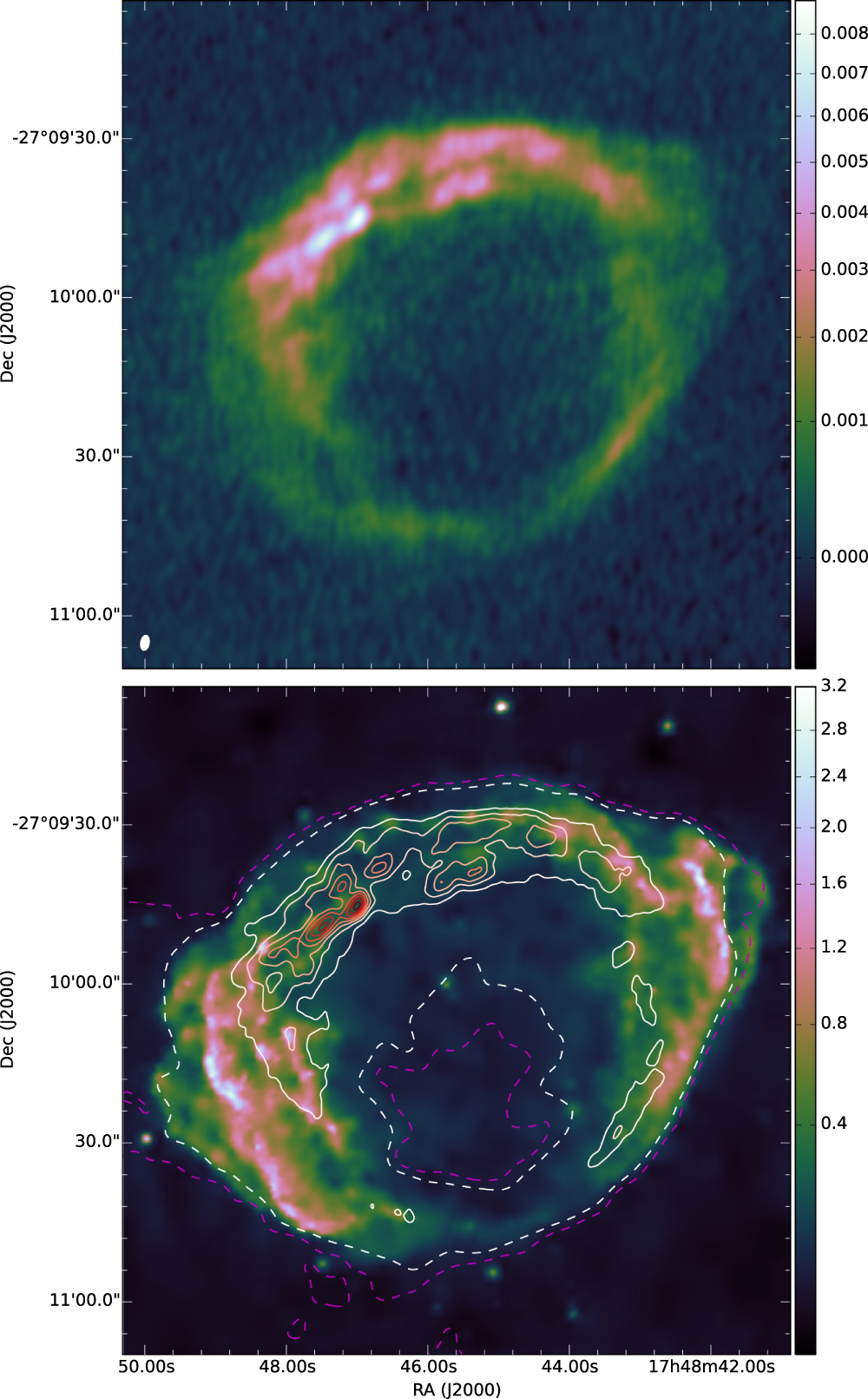}
    \caption{Top: Total intensity VLA image of G1.9+0.3 at 1365 MHz. The resolution is $2.^{\prime\prime}8\times1.^{\prime\prime}6$ at a PA of -9.$^{\circ}$6 (as shown in the left corner). The scale is in mJy beam$^{-1}$. Bottom: Smoothed 2009 1.2-8 keV \textit{Chandra} image overlaid with selected radio contours emphasising bright (solid lines from 1 to 8 mJy~beam$^{-1}$ spaced by 1 mJy~beam$^{-1}$) and very faint (dashed lines in magenta and white at 0.06 and 0.12 mJy beam$^{-1}$) emission. The scale is in counts per $0.^{\prime\prime}246\times0.^{\prime\prime}246$ image pixel (half an ACIS pixel). Intensities are shown with the cube helix colour scheme of \citet{Green2011}. This figure was reproduced by permission of the AAS from \citet{borkowski2017}.}
    \label{fig:Borkowski}
\end{figure}

The SNR G1.9+0.3 is the youngest galactic remnant \citep{reynolds2008}, first discovered by \citet{green1984}. The remnant's age ranges anywhere between 100 and 150 years \citep{borkowski2017,luken2020}. Observational studies indicate that the progenitor star probably exploded as a type Ia supernova \citep{borkowski2017}. 
Its mean angular size is 1.5 arcminutes. Considering a distance of 8.5 kpc, based on X-ray absorption, HI,  and molecular emission studies \citep{reynolds2008,luken2020} results in a radius of 1.8 pc.
As shown by \citet{luken2020}, this object exhibits an incomplete shell morphology in the radio continuum, increasing in brightness to the north, with a spectral index average of -0.6. However, in X-ray observations this remnant appears elongated \citep[see][]{reynolds2008, borkowski2017}, with two bright arcs or ear-shaped structures to the southeast and northwest. This striking difference between the non-thermal radio and X-ray brightness distributions 
is one of the most intriguing features of this remnant. Furthermore, both emissions have a non-thermal origin. An image of both the radio and X-ray emission can be seen in Figure~\ref{fig:Borkowski}, reproduced with permission of the AAS from \cite{borkowski2017}.\\

Some studies \citep{reynolds2008,tsuji2021} suggest that the synchrotron mechanism is responsible for the observed X-ray emission. Those studies also showed that a synchrotron model with an exponential energy cut-off fits well with the X-ray spectrum of this remnant.
Nevertheless, the radio and X-ray emissions produced by the synchrotron process display practically identical brightness distribution in the SN 1006 case \citep{winkler2014}. Based on this last fact, \citet{borkowski2017} proposed searching for a different mechanism to reconcile these differences for the case of SNR G1.9+0.3. For example in \cite{2019A&A...627A.166B}, the authors explored a scenario where the synchrotron radio and the synchrotron X-ray emissions come from different shocks in order to generate the spatial discrepancies between emissions.

Observational studies of the expansion of SNR G1.9+0.3 \citep{carlton2011,borkowski2014,borkowski2017,luken2020}, revealed different expansion velocities between the northern part and the "ears". \citet{borkowski2017} suggested that the remnant's northern part has collided with a dense circumstellar shell region and that the propagation of the supernova blastwave is aspherical.
Based on this expansion study and doing some estimates of the maximum energy achieved by accelerated electrons, \citet{borkowski2017} explain the lack of non-thermal X-ray emission towards the north region of this remnant due to the low expansion velocity in this region.
Recently in \cite{Enokiya2023}, through the use of line observations in $^{12}$CO and $^{13}$CO, the authors found a molecular cloud in the immediacy of G1.9+0.3. This further supports the idea of SNR expanding into an in-homogeneous media. At higher energies, in \cite{2014MNRAS.441..790H} the authors searched for $\gamma$-ray emission using the H.E.S.S. Cherenkov telescope array, however they were unable to detect a significant signal from G1.9+0.3.

Several theoretical studies try to explain this astrophysical object's distinctive morphology, emission, and expansion. For example, \citet{tsebrenko15b} analysed the expansion of an SNR inside a planetary nebula. 

Following the idea given by \citet{borkowski2017}, \citet{zhang2023} carried out adiabatic 3D MHD simulations, where they consider this object an SNR that expands into an interstellar medium until it collides with a dense cloud. From their numerical results, these authors performed synthetic radio and X-ray emission maps obtaining spherical morphologies, i.e. they do not produce the characteristic ears of SNR G1.9+0.3.
Recently,  \citet{Soker2023} takes a step further, finding a large-scale point-symmetry and indicating that explaining G1.9+0.3 requires the explosion of a SN Ia into a planetary nebula.

Considering some of the hypotheses adopted in previous studies \citep[e.g.,][]{tsebrenko15b,zhang2023,velazquez2023}, we have carried out 3D MHD numerical simulations using the  code {\sc guacho} \citep{Esquivel2009,Villarreal2018}, endeavouring to characterise the morphology and emission of this puzzling astrophysical object. Instead of invoking the synchrotron mechanism to explain the X-ray non-thermal emission, we propose a possible alternative which does not require a strong magnetic field: Inverse Compton (IC) radiation, which could achieve geometrical distributions similar to those observed with CHANDRA and with NuSTAR \citep[]{2015ApJ...798...98Z} while maintaining a non-thermal energy distribution. This idea has been explored analogously, in \cite{2023MNRAS.524.4939M} the authors observed a spatial discrepancy between the non-thermal radio and X-ray emission in a group of galaxies at z = 0.131, attributing the radio photons to synchrotron and the X-ray ones to inverse Compton processes.

We organise the present work as follows:  Section \ref{Sec:2} describes the characteristics of the studied scenarios and the initial setup of the simulations. The results of our modelling and their analysis are presented in Section \ref{Sec:3}. Finally, we summarise our main conclusions in Section \ref{Sec:4}.

\section{Simulations: initial setup}\label{Sec:2}

This Section presents the scenarios explored in this study, the used numerical methods, the manner in which synthetic emission maps are generated and the model developed for the Inverse 
Compton emission. 

\subsection{The scenario}

\begin{figure}
    \centering
\includegraphics[width=\columnwidth]{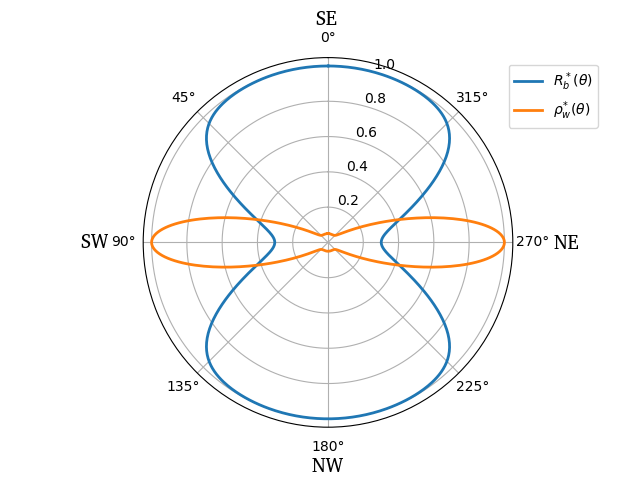} 
    \caption{Polar dependence of the normalised wind bubble density distribution $\rho^*_{\rm w}(\theta)=\rho_{\rm w}(\theta)/\rho_{\rm {w,max}}$ (orange line)  and  radius $R^*_{\rm b}(\theta)=~R_{\rm b}(\theta)/ R_{\rm b, max}$ (blue line), obtained for $\alpha=0.95$ and $\beta=5$.}
    \label{fig:rbw_rhow}
\end{figure}

Our goal is to model the morphology and non-thermal emission of SNR G1.9+0.3. This object exhibits an elliptical morphology, whose large axis is along the SE-NW direction, with a mean radius of 1.5 pc. At radio frequencies, this object looks like an incomplete shell, with high emission towards the NE. However, the non-thermal X-rays emission concentrates in the SE and NW regions outside the synchrotron shell.

\citet{zhang2023} presented 3D MHD simulations for modelling this object. They considered an SNR expanding into an ISM with a density of 0.21 $\rm cm^{-3}$ and an interstellar magnetic field of 1~$\mu$G. To emulate the increase in synchrotron emission towards the NE of the remnant, these authors assume that the SNR collides with a cloud \citep[following the idea given by][]{borkowski2017}. Furthermore, they also computed synthetic non-thermal emission maps obtaining a radio emission increase in the SNR-cloud collision zone and regions with high X-ray emission. However, their models do not produce an elongated SNR shell.

Several authors reported that elongated `ear-like' SNR morphologies result from the interaction of a supernova ejecta with an anisotropic circumstellar medium produced by the progenitor or companion stellar wind \citep{blondin1996,2020Galax...8...38C, alexandros21, ustamujic2021,meyer_2022,velazquez2023}. In particular, \citet{velazquez2023} explored the formation of elongated SNRs due to the interaction of the SNR shock front with a circumstellar medium, which has a dense and narrow equatorial region. Their models with a narrow equatorial region produce peanut-like stellar wind bubbles. Our initial guess is that this configuration would produce the observed synchrotron emission of G1.9+0.3 if we also consider an interstellar magnetic field almost parallel to the stellar wind bubble axis. In this way, we expect stronger radio emissions  in equatorial latitudes than in the polar regions because, in the former regions, the SNR shock wave is sweeping up the ISM magnetic field. In contrast, the shock front does not directly interact with the ISM in the polar zones. 

Bipolar circumstellar structures with a density enhancement at the equatorial plane frequently surround stars with dense and slow stellar winds, such as stars at the Asymptotic Giant Branch \citep[AGB;][]{decin2020}. This kind of bipolar structure results from angular momentum transport between the stellar parent system's orbital motion and its stellar winds
\citep[e.g.][]{Bjorkman1993, Mastrodemos1999, Heger2000, Politano2011, gloria2020}.

We carried out our numerical simulations in two phases based on these previous works. In the first one, we modelled a stellar wind bubble's evolution and formation; in the second one, we introduced the SN ejecta in the center of the wind bubble within 
it, see also methods in \citet{meyer2020,meyer2023}. 
We considered two scenarios or runs that are distinguished by the density profile imposed to the ISM. In the first run (R1), we assume a constant number density $n_0=0.25\, \mathrm{cm}^{-3}$, a temperature of 1000 K, and a magnetic field $B_0=1 \mathrm{ \mu G}$~\citep{zhang2023}. In the second (R2) run, we considered the idea given by \citet{borkowski2017} that the remnant is interacting in the north with a dense region an obstructed-expansion model. Unlike \citet{zhang2023}'s work, where they imposed a medium with a strong density contrast, we explore the case that the interstellar medium has a density with an exponential profile, given by:
\begin{equation}
    n=n_0\exp{(-d/H)}
\end{equation}
with $d$ being the distance from the computational domain centre along the direction ( 0 , $-\sin~30^\circ$ , $\cos~30^\circ$ ), and $H=2.5$~pc is the characteristic ISM increasing density length.

To describe the stellar wind density, we followed the equations of \citet{mellema1991} that provide the density distribution as a function of radius $r$ and polar angle $\theta$ in spherical coordinates: 
\begin{equation}
\rho_w(\theta, r)=\frac{\dot{M}_{\rm w}}{4\pi v_{\rm p} r^2}f(\theta),
    \label{eq:rhow}
\end{equation}
where $\dot{M}_{\rm w}$ is the mass-loss rate and $v_{\rm p}$ is the terminal velocity of the stellar wind at the pole ($\theta= 0$). We have set $\dot{M}_{\rm w}=3\times 10^{-6} \textrm{M}_{\odot}\, \textrm{yr}^{-1}$ and $v_p=20\, \textrm{km}\,\textrm{s}^{-1}$, which are typical values for AGB stars \citep[e.g.][]{Vassiliadis1993}. The function $f(\theta)$ is given by:
\begin{equation}
    f(\theta)=\frac{1}{1-\alpha}\left[1-\alpha\frac{1-\exp{(-2\beta \cos^2\theta)}}{1-\exp{(-2\beta)}}\right],
    \label{eq:ftheta}
\end{equation}
where $0 \le \alpha < 1$ gives the equator-to-pole density ratio which (given by $(1-\alpha)^{-1}$, and $\beta$ determines the width of the equatorial region ($\theta\sim \pi/2$), i.e.  $\beta>1$ ($\beta<1$) produces a narrow (wide) equatorial region. 

The wind velocity distribution also depends on the polar angle as:
\begin{equation}
    v_{\rm w}(\theta)=\frac{v_{\rm p}}{f(\theta)}.
    \label{eq:vel}
\end{equation}
In this way, the mass loss rate $\dot{M}_w$ is isotropic.

We chose $\alpha=0.95$, which implies an equator-to-pole density ratio of 20, and $\beta=5$ \citep{meyerwr2021,Meyer2021,velazquez2023}. In Figure~\ref{fig:rbw_rhow} $\rho^*_{\rm w}(\theta)=\rho_{\rm w}(\theta)/\rho_{\rm {w,max}}$ is plotted as the orange line, a thin horizontal structure, for the for the chosen values of parameters $\alpha$ and $\beta$.
The normalised radius $R^*_{\rm b}(\theta)$ of the stellar bubble is given by \citep{alexandros21}: 

\begin{equation}
    R^{*}_{\rm b}=\frac{R_{\rm b}(\theta)}{R_\textrm{b,max}}=[f(\theta)]^{-2/5}
    \label{eq:rb},
\end{equation}
being $R_\textrm{b,max}=R_b(\theta=0)$ the radius of the  wind bubble at the pole. Figure \ref{fig:rbw_rhow} displays $R^*_{\rm b}(\theta)$, as the blue contour, showing a vertical peanut-like shape.

In the first phase of the simulations, we imposed the stellar wind condition as an interior boundary condition, given by a spherical surface with radius $R_{\rm w}=0.12$~pc (centred in the middle of the computational domain). We tilted the polar axis of the stellar wind (which is contained in the $x=0$ plane) by 60$^{\circ}$ with respect to the $z$-axis so that the polar direction of the wind does not coincide with any of the simulation axes to avoid possible numerical artefacts stemming from the Cartesian grid.
The ambient magnetic field is on the $xy$ plane, making an angle of 30$^{\circ}$ with the $x$-axis. We let both runs evolve for 150~kyr to form the wind bubble \citep{hofner2018,Hernandez2019}.

Once the stellar wind bubble formed, in the second phase we imposed a type Ia SN explosion in a sphere of radius $R_0=0.16$~pc at the centre of the computational domain. The initial energy was set $E_0=10^{51}$~erg \citep{martinez2022}. A fraction $f_{\rm K}=0.95$ of $E_0$ corresponds to the kinetic energy which can be transformed into other forms of energy, such as magnetic energy or radiation. The mass $M_{*}$ ejected by the Type Ia SN was set as 1.38$\mathrm{M}_\odot$ we can estimate an SNR initial age of $\simeq 10$~yr \citep{truelove1999}. Furthermore, we considered that at this radius, the SNR swept up a CSM mass of $0.1\textrm{M}_\odot$
which was added to the initial SN mass.

The initial remnant has a constant density $\rho_{\rm c}$ from the centre up to a radius $r_{\rm c}$, while for $r_{\rm c}\leq r\leq R_0$, the density is $\rho_{\rm c} (r_{\rm c}/r)^7$ \citep{Jun1996}. The outer region contains a fraction $X_{\rm m}$ of $M_0$ while the remaining $(1-X_{\rm m})M_0$ was uniformly distributed in the inner sphere with radius $r_{\rm c}$. Furthermore, the radial velocity $v_r$ grows linearly with $r$ as $v_r=v_0 (r/R_0)$, $v_0$ being the velocity at $r=R_0$. The magnitudes $r_{\rm c}$, $\rho_{\rm c}$, and $v_0$ are given by \citep{velazquez2023}:
\begin{eqnarray}
  r_c&=&R_0 \bigg[\frac{1-(7/3)X_m}{1-X_m}\bigg]^{1/4}, \\
  \rho_c&=&\frac{3 M_0}{4\pi R^3_0}\frac{(1-X_m)^{7/4}}{(1-(7/3) X_m)^{3/4}},\ \text{and}\\
  v_0 &=& \sqrt{\frac{4 f_k E_0}{3 M_0 (1-X_m) y^2_r(7/5-y^2_r)}},
\end{eqnarray}
with $y_r=r_{\rm c}/R_0$. We set $X_m=0.4$.

\subsection{The code}

The numerical study was performed with the parallel 3D magneto-hydrodynamical (MHD) code {\sc guacho} \citep{Esquivel2009,Villarreal2018}. This code solves the ideal MHD equations in a fixed Cartesian grid :

\begin{equation}
    \frac{\partial\rho}{\partial t}+\nabla\cdot(\rho\mathbfit{u})=0\;,
	\label{eq:mass}
\end{equation}

\begin{equation}
    \frac{\partial(\rho\mathbfit{u})}{\partial t}+\nabla\cdot\left[\rho\mathbfit{u}\otimes\mathbfit{u}+\mathbfss{I}\left(p+\frac{B^2}{8\pi}\right)-\frac{\mathbfit{B}\otimes\mathbfit{B}}{4\pi}\right]=0\;,
	\label{eq:momentum}
\end{equation}

\begin{equation}
    \frac{\partial e}{\partial t}+\nabla\cdot \left[
    \left(e + p + \frac{B^2}{4\pi}\right)\mathbfit{u}-\left(\mathbfit{u} \cdot \mathbfit{B}\right) \mathbfit{B}
    \right]=Q_L\;,
	\label{eq:energy}
\end{equation}

\begin{equation}
    \frac{\partial \mathbfit{B}}{\partial t}-\nabla\times \left(\mathbfit{u}\times \mathbfit{B}\right)=0\;,
   	\label{eq:induction}
\end{equation}

\noindent where $\rho$, $\mathbfit{u}$, $p$, $\mathbfit{B}$ and $e$ are the mass density, velocity, gas pressure, magnetic field and total energy density, respectively. In Eq. (\ref{eq:momentum}), $\mathbfss{I}$ is the identity matrix. The energy density is given by $e=\rho u^2/2+p/(\gamma-1)+B^2/8\pi$, with $\gamma$ being the heat capacity ratio of the gas, which was set to 5/3. The code includes radiative cooling $Q_L=n^2 \Lambda(T)$ (see Eq. \ref{eq:energy}), where $n$ is the gas number density and $\Lambda(T)$ is a parameterised function of the temperature, which describes an optically-thin cooling given by \citet{Dalgarno1972}. A second-order Godunov method with the approximate Riemann solver HLLD \citep{miyoshi2005}, was used to advance Eqs. (\ref{eq:mass})--(\ref{eq:induction}) in time. A zero-gradient (outflow) condition is imposed in all of the domain boundaries.

\subsection{Synthetic emission maps}

The synthetic observations created for this work required quantities in multiple frames of reference. The first one is an in situ one (xyz frame) while the second one is placed in the plane of the sky (x'y' plane) and the line of sight (LoS, z' coordinate). The second frame can be rotated along the LoS to make our synthetic maps visually similar to the familiar form of the SNR G1.9+0.3.

\subsubsection{Synchrotron  emission}
\label{section:Synchrotron_equations}

Synchrotron emission is produced by the interplay between two key ingredients: a magnetic field and relativistic particles. Producing synthetic synchrotron emission maps requires that in each computational cell, there is an assigned relativistic electron population and a known magnetic field. 

To estimate the first quantity, we used the following density distribution of electrons 
\begin{equation}
    N_e(\gamma)=K \gamma^{-p},
    \label{eq:density_distro_electron}
\end{equation}
where $K$ is a constant, $\gamma$ is the electron Lorentz factor and the $p$ index is related to the spectral index $\alpha$ by:

\begin{equation}
\alpha = \frac{p-1}{2}.
\label{eq:SpectralIndex}
\end{equation}

A connection with the physical data produced in our simulations must be established. This was achieved by taking into account that the electron density and energy can be related to the number density and energy density of the gas ($n_g$ and $\epsilon_g$, respectively) by means of:
\begin{equation}
\chi_n n_g = n_e = \int_{\gamma_{min}}^{\infty}K \gamma^{-p} d\gamma \simeq \frac{K}{p-1} \gamma_{min}^{-p+1}
\label{eq:chi_n}
\end{equation}
\begin{equation}
\begin{aligned}
\chi_\epsilon \epsilon_g =  \int_{\gamma_{min}}^{\infty}K \gamma^{-p} (\gamma - 1) m_e c^2 d\gamma \simeq \\
\simeq m_e c^2 \frac{K}{p-1}\gamma_{min}^{-p+1} \bigg(\frac{p-1}{p-2} \gamma_{min}-1\bigg), 
\label{eq:chi_e}
\end{aligned}
\end{equation}
with $\chi_n$ and $\chi_\epsilon$ being the gas density and energy fractions converted in the electron density and energy. Utilizing those equations and considering that:
\begin{equation}
    \dfrac{(p-1)}{(p-2)} \gamma_{min} \gg 1,
\end{equation}
 the values of K and $\gamma_{min}$ can be written as:
\begin{eqnarray}
    K&=& (p-1) \chi_n n_g \gamma_{min}^{p-1} \label{eq:K}\\
\gamma_{min}&=&\frac{\chi_e\epsilon_g (p-2)}{\chi_n n_g (p-1)m_e c^2}. 
\label{eq:gmin}
\end{eqnarray}

For the value of the gas' energy density, we used the thermal energy density, $\epsilon_g = c_v P_g$, c$_v$ being the specific heat and P$_g$ the pressure of the gas. This completes establishing a bridge between our simulation's data and the relativistic particle population. 

We next employ Eq 4.43 of \cite{2013LNP873G} to calculate the synchrotron emissivity $j_s(\nu,\theta)$ :
\begin{equation}
    j_s(\nu,\theta)=\frac{3 \sigma_T c K U_B}{8 \pi^{2} \nu_L} (\sin\theta)^{(p+1)/2}\bigg(\frac{\nu}{\nu_L}\bigg)^{-(p-1)/2} f_s(p),
    \label{eq:syn_emi}
\end{equation}
where $K$ is the constant given by Eq.\ref{eq:K}, $\sigma_T$ is the Thomson cross section, $c$ is the speed of light, $U_B=B^2/(8\pi)$ is the magnetic energy density, $\nu_L=(e B)/(2\pi m_e c)$ is the Larmor's frequency, $\theta$ is the pitch angle between the relativistic electrons and the magnetic field and $f_s(p)$ is a function that depends on $p$ index:
\begin{equation}
    f_s(p)=3^{p/2}\frac{\Gamma(\frac{3p-1}{12})\Gamma(\frac{3p+19}{12})}{p+1}. 
    \label{eq:f_s}
\end{equation}

Considering that the detected synchrotron photons should originate from electrons gyrating along magnetic lines not parallel to the line of sight, from our available information, we selected the magnetic components residing in the planes perpendicular to the LoS. Substituting $B_\perp = B sin \theta$ into Eq. \ref{eq:syn_emi}, both in $U_B$ and in $\nu_L$, together with the value of $K$, we calculated the emissivity in each cell with the following equation:
\begin{equation}
\begin{aligned}
        j_s(\nu)=3\sigma_T \sqrt{\frac{e^{p-3} c^{9-5p}}{2^{p+9} \pi^{p+3} m_e^{3p-5}}} \cdot \\
    \cdot \frac{[\chi_\epsilon c_v (p-2)]^{p-1}}{[\chi_n (p-1)]^{p-2}} \cdot \\ 
    \cdot \frac{P_g^{p-1}B_{\perp}^{(p+1)/2}}{n_g^{p-2}} \nu^{-(p-1)/2} f_s(p).
\end{aligned}     
\end{equation}

To obtain synthetic synchrotron emission maps, we computed the total intensity of the synchrotron emission by integrating the emissivity along the line of sight:  
\begin{equation}
I(x^\prime,y^\prime,\nu)=\int_\textrm{LoS}j_{\rm s}(x^\prime,y^\prime,z^\prime,\nu)dz^\prime, 
\label{eq:I}
\end{equation}  

The Stokes parameters $Q$ and $U$ are obtained from the specific emissivity as:
\begin{equation}
Q(x^\prime,y^\prime,\nu)=\int_\textrm{LoS}f_{\rm p}\,j_{\rm s}(x^\prime,y^\prime,z^\prime,\nu)\cos(2\phi)dz^\prime,
\label{eq:Q}
\end{equation}  
\begin{equation}
U(x^\prime,y^\prime,\nu)=\int_\textrm{LoS}f_{\rm p}\,j_{\rm s}(x^\prime,y^\prime,z^\prime,\nu)\sin(2\phi)dz^\prime,
\label{eq:U}
\end{equation}
with $\phi$ the position angle of the local magnetic field in the plane of the sky, and  $f_{\rm p}$ is the linear polarization degree, related to the spectral index $\alpha$, viz:
\begin{equation}
f_{\rm p}=\frac{\alpha+1}{\alpha+5/3}.
\end{equation}  

Finally, the magnetic field position angle ($\Phi_B$) is obtained by, 
\begin{equation}
    \Phi_B(x^\prime,y^\prime,\nu)=\frac{1}{2}\arctan\bigg( \frac{U(x^\prime,y^\prime,\nu)}{Q(x^\prime,y^\prime,\nu)}\bigg)
\end{equation}

\subsubsection{Inverse Compton emission}

As mentioned before, one of the possible sources for the X-ray emission detected is that created via inverse Compton processes. A distribution of relativistic particles that encounters a radiation field could trigger IC emission. The electrons responsible for the synchrotron emission could start the IC effect in the medium. Hence, we considered the same electron distribution from eq. \ref{eq:density_distro_electron}. When the aforementioned electron population encounters an isotropic field of monochromatic photons, it results in an emissivity described by equation 5.51 of \cite{2013LNP873G}, that is:
\begin{equation}
\epsilon(\nu_c) = \dfrac{1}{4\pi} \dfrac{(4/3)^\alpha}{2} \sigma_T c K \dfrac{U_r}{\nu_0} \bigg( \dfrac{\nu_c}{\nu_0} \bigg)^{-\alpha},
\label{eq:EmissivityIC}
\end{equation}
where the emissivity $\epsilon$ is observed at a frequency $\nu_c$ depends on the product of the Thompson cross-section $\sigma_T$, the speed of light c, the constant K described in eq. \ref{eq:gmin} and an energy density of monochromatic radiation $U_r$  from photons with a frequency $\nu_0$. The quotient of frequencies is elevated to the spectral index $\alpha$. We note that the IC emissivity has the same functional dependence on frequency as the synchrotron emissivity (see Eq. \ref{eq:syn_emi}). A log-log graph of flux vs frequency would have the same slope, which would not allow us to discern between these emission processes reliably.

The energy density $U_r$ can be estimated by: 
\begin{equation}
U_r = \dfrac{L}{V} t_{\text{esc}},
\label{eq:IC_EnergyDensity}
\end{equation}
in which the energy density originates in a source of volume $V$ with a luminosity $L$ where the photons take on average a time $t_{esc}$ to escape it. The escape time can be rewritten considering that the photon travels a distance $R$,
\begin{equation}
t_{\text{esc}} = \dfrac{3R}{4c}.
\label{eq:EscapeTime}
\end{equation}

To proceed, we considered each cell of our simulation as an isotropic source of length $2R$ for a monochromatic set of photons. 

A star's wind bubble can be observed in the optical range, in part, due to the $\rm H\alpha$ emission \citep[see][]{giovanardi1989}. This monochromatic line emission has been well characterized throughout the years, and it can be confidently estimated in each cell of our MHD simulation. For this reason, we used the estimated $\rm H\alpha$ emissivity as an energy source. Substituting eq. \ref{eq:EscapeTime} in eq. \ref{eq:IC_EnergyDensity} and remembering that an emissivity at a given frequency is the luminosity per volume per unit of solid angle, we rewrite $U_r$ as:
\begin{equation}
U_r = \dfrac{3}{4} \dfrac{R}{c} 4\pi \epsilon_{H\alpha}(\nu_{H\alpha}),
\label{eq:HA_EnergyDensity}
\end{equation}
where $\nu_{H\alpha} \sim 4.57 \times 10^{14}$ Hz is the H$\alpha$ transition frequency. 

Using this method, it is possible to adapt different radiation sources as the base of the energy density $U_r$. 
Due to the nature of our simulation, a known radiation source is the radiative cooling $Q_L$ (see Eq.\ref{eq:energy}) which can be used as a proxy of background radiation considering that each cell in our simulation has a corresponding parameterized cooling value at each time step. In this case, the density energy is given by:
\begin{equation}
    U_r=\dfrac{3}{4} \dfrac{R}{c} Q_L
    \label{eq:QL_EnergyDensity}
\end{equation}
Finally, we can integrate the obtained emissivity (Equations \ref{eq:EmissivityIC}, \ref{eq:HA_EnergyDensity}, and \ref{eq:QL_EnergyDensity}) along the line of sight to obtain intensity maps. 

\section{Results}\label{Sec:3}

This Section presents the evolution of the supernova remnants, describes the simulated emission maps 
and compare them with available observations. 

\subsection{Distribution of physical quantities}

The setups used for the simulations in the present work result in two key moments. The first one at 150 kyr, after the wind has plowed the vicinity of the original stars but before the SNR begins expanding. The second moment is 140 years after the first one. This is roughly similar to the expected age of G1.9+0.3 and accommodates an SNR with a radius $\sim 2$ pc.

\begin{figure*}
    \centering
    \includegraphics[width=16cm]{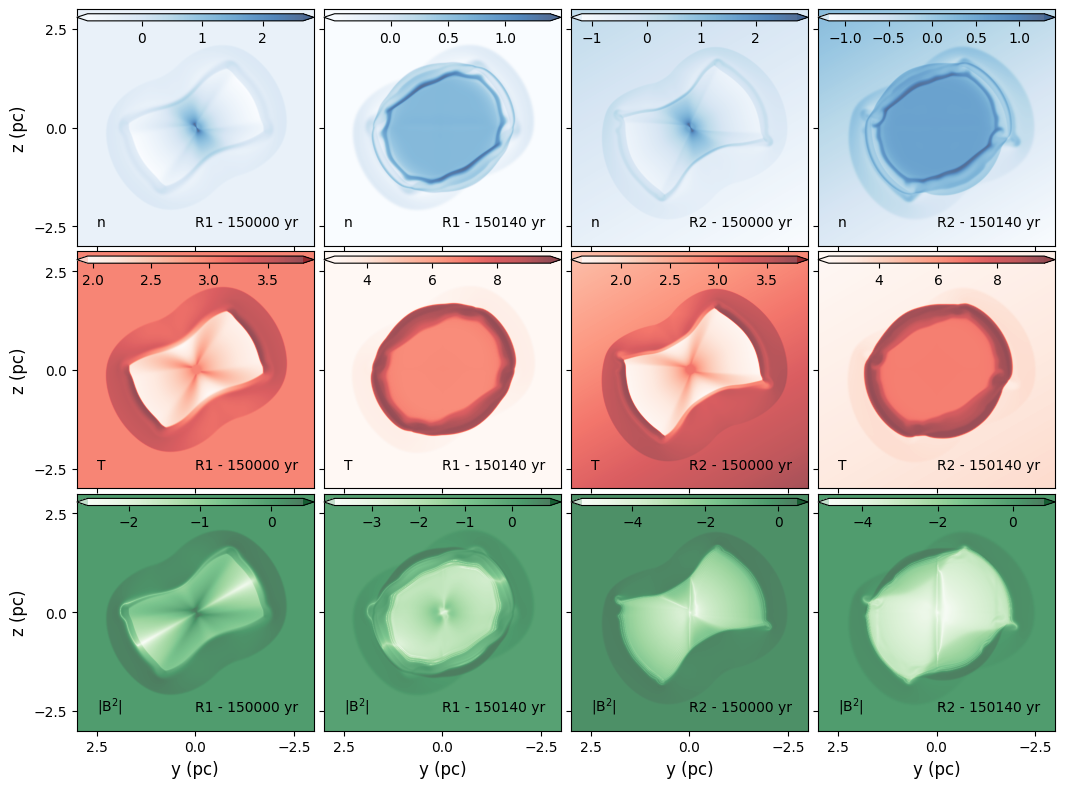}
    \caption{Slices of the MHD simulations along the YZ plane showing logarithmic distributions of density (blue) [cm$^{-3}$], temperature (red) [K] and magnetic field intensity (green) [$\mu$G], for the R1 (first and second columns) and the R2 models (third and fourth columns). The earliest time (150000 yr) is taken before the SN explosion while the latest (150140 yr) includes the expanding SNR.}
    \label{fig:MHD_maps}
\end{figure*}

The initial medium, be it uniform or with the density gradient used, suffers transformations from its interaction with the pre-supernova wind, and later, the SNR passes through it. The final distributions, after 150 kyr and after 150140 yr, are shown in Figure \ref{fig:MHD_maps}, with the top being density (blue shades), the middle row being temperature (red shades), and the bottom row corresponding to magnetic field intensity (green shades). The image is divided into four columns. The first two correspond to the simulation with uniform initial conditions, called henceforth the R1 model, and the third and fourth columns correspond to what we denominate the R2 model, which has a density gradient in its initial conditions.

From the image, it's clear that the wind changes the dynamics of the expanding SNR. The shock front appears smoother in regions depleted by the wind. The expansion of the remnant in conjunction with the wind's particular geometry, results in a shock that encounters the ISM at different times, creating regions of high density closer to the origin of the explosion in the north-west and south-east. This creates an object with a box-like centre and two prominent limbs at the sides. The temperature distribution doesn't seem to have a correlation with the density one, i.e. the hotter areas don't correspond with the densest ones. The magnetic field intensity distribution appears to have a geometry similar to the density. 

Finally, it is worth noting that the SNR in the obstructed expansion model breaks the north-west/south-east symmetry in the three quantities shown. This is the main difference with the free-expansion model. 

\subsection{Synthetic emission maps}
To obtain the synthetic maps, the computational domain was rotated -90$^\circ$ in $\hat{x}$, -120$^\circ$ in $\hat{y}$, and -60$^\circ$ in $\hat{z}$.

\subsubsection{Synchrotron radio maps}

\begin{figure*}
    \centering
    \includegraphics[width=16cm]{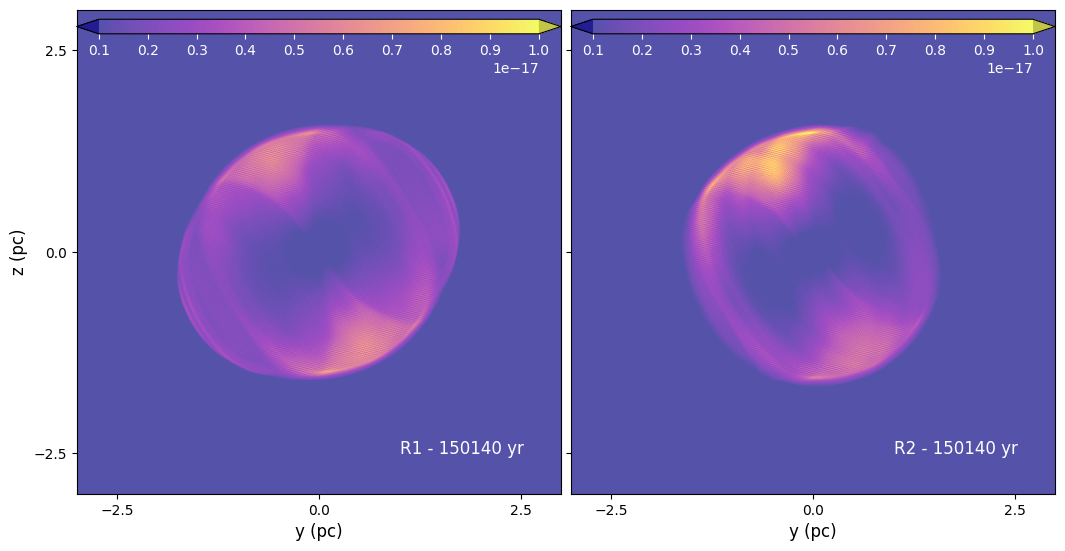}
    \caption{Synchrotron emission maps in erg~cm$^{-2}$~sr$^{-1}$~s$^{-1}$ Hz$^{-1}$ at 2.1 GHz, obtained from the simulation data for both the R1 (left) and R2 (right) models.}
    \label{fig:Syncro_maps_raw}
\end{figure*}

Following the procedure described in section \ref{section:Synchrotron_equations} we constructed maps of the synchrotron emission artificially observed at a radio frequency of 2.1 GHz. The images presented in Figure \ref{fig:Syncro_maps_raw} correspond to the emissivity of the SNR after 140 yr of expansion in erg~cm$^{-2}$~s$^{-1}$~sr$^{-1}$ observed as if the beam was the size of each pixel. The one on the left corresponds to R1, the free-expansion model, while the one on the right to R2, the obstructed-expansion one. While both objects have peaks around $10^{-17}$ [erg~cm$^{-2}$~s$^{-1}$~sr$^{-1}$] the differences between them are clear. The R1 model has a box-like structure with limbs at the sides, and the brightest areas are situated in the NE and SW sectors of the SNR. In contrast, the R2 model doesn't have apparent limbs, and the NE area is prominently brighter than the SW one.        

\begin{figure*}
    \centering
    \includegraphics[width=16cm]{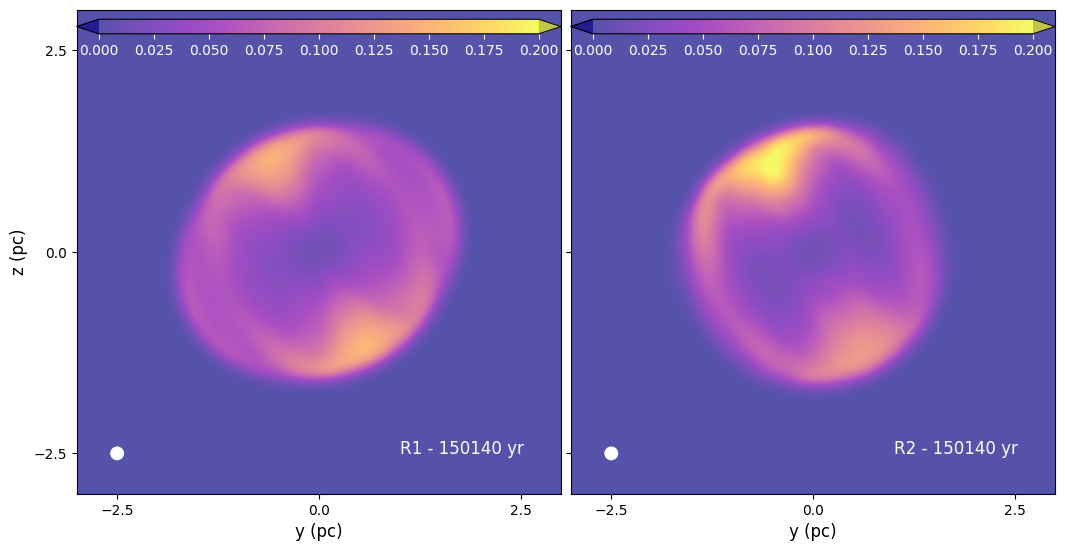}
    \caption{Synchrotron emission maps in mJy~beam$^{-1}$ at 2.1 GHz, obtained from the simulated data after applying a smoothing filter, for the R1 (left) and R2 (models). The virtual beam used for the synthetic observation has a radius $\sim 2^{\prime\prime}$ and is represented by an equivalent circle in the lower left corner. }
    \label{fig:Syncro_maps_smt}
\end{figure*}

In order to facilitate the comparison of our data with real observations, we applied a smoothing Gaussian filter with a $\sigma = 6.5$ pixels. This technique in conjunction with assuming that the object is situated at a distance $d = 8.5$ kpc from the observer and that the observation has a beam of radius $\sim 2"$ were used to create the maps shown in Figure~\ref{fig:Syncro_maps_smt}. There it is clear that while the geometrical characteristics of the observed object are the same as those from the clean image, the value of the emissivity peaks now has a value significantly larger, being close to $j \sim 0.2$ mJy~beam$^{-1}$. 

\begin{figure*}
    \centering
    \includegraphics[width=16cm]{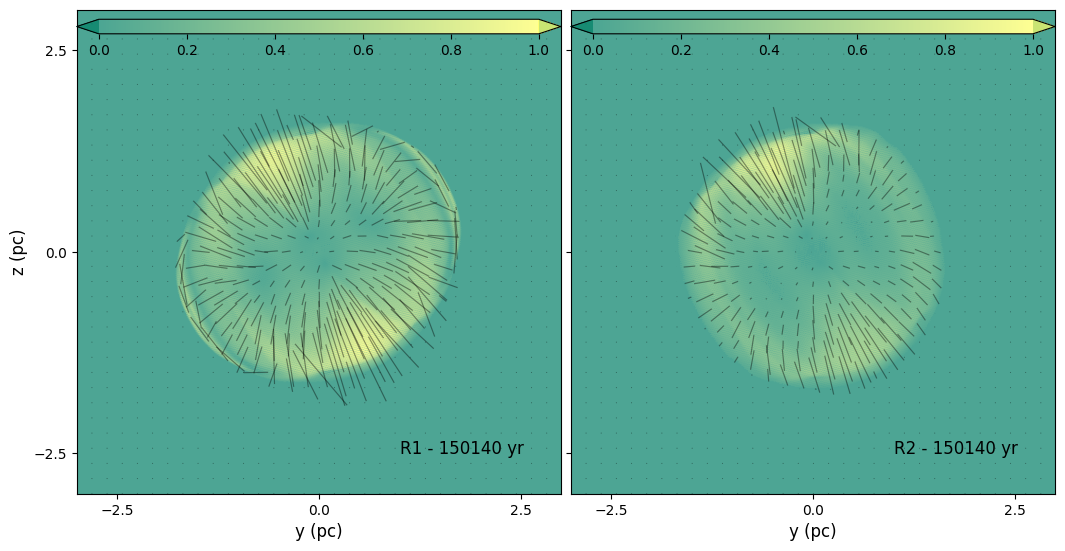}
    \caption{Polarized emission maps with the magnetic field vectors (black) from the Stoke parameters Q and U for both the R1 (left) and R2 (right) models.}
    \label{fig:Polarization_maps_QU}
\end{figure*}

Throughout the use of the Stoke parameters Q and U we constructed a map of the magnetic field in the selected plane of observation and additionally a normalized distribution of the linearly synchrotron emission was added as a background; these results are presented in Figure \ref{fig:Polarization_maps_QU}. The magnetic field vectors on the simulated plane of the sky from the polarized emission maps show a radial distribution in general. However, the left plot, corresponding to R1, has magnetic vectors with a visible change in direction at the regions corresponding to the lateral shocks.

\subsubsection{Inverse Compton non-thermal X-ray maps}

\begin{figure*}
    \centering
    \includegraphics[width=16cm]{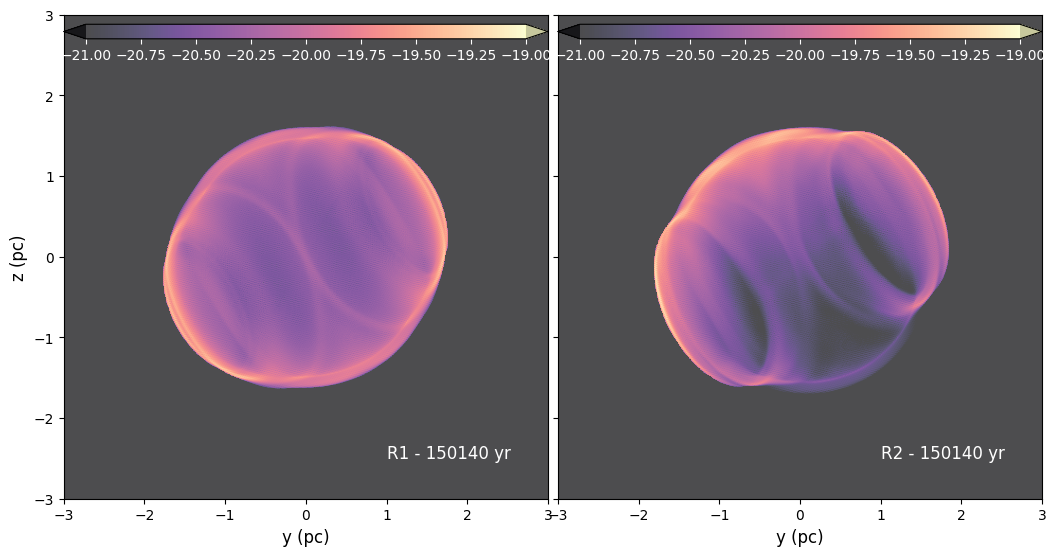}
    \caption{X-ray inverse Compton logarithm intensity maps in erg s$^{-1}$ cm$^{-2}$ sr$^{-1}$ for a 2-10 KeV band where the background radiation comes from H$\alpha$ emission. The R1 model is on the left while the R2 model is on the right.}
    \label{fig:InvCom_maps_raw}
\end{figure*}

After building a map of the H$\alpha$ emission and following the procedure described before, in Figure \ref{fig:InvCom_maps_raw} we illustrate the synthetic X-ray observations of the inverse Compton radiation due to the aforementioned H$\alpha$ energy field. The plots were constructed by selecting an observation band constrained between 2 and 10 keV. The left image corresponds to the R1 model; it entirely outlines the SNR and at the same time, it shows an important intensity difference between the NE-SW and the NW-SE regions regions, i.e., the sides are brighter than the caps. The brightest regions show an intensity $j_{IC} \sim 10^{-19}$ erg s$^{-1}$ cm$^{-1}$ sr$^{-1}$ while the caps have an intensity $j_{IC} \sim 10^{-20}$ erg s$^{-1}$ cm$^{-1}$ sr$^{-1}$, almost one order of magnitude smaller. The right image, corresponding to R2, is not a complete outline of the object. The SW region basically disappears. Additionally, the difference in intensity between the sides and the top cap is harder to declare. The NE region of the R2 model has non-negligible IC emission with both the sides and the visible cap having points of intensity $j_{IC} \sim 10^{-19}$ erg s$^{-1}$ cm$^{-1}$ sr$^{-1}$. 

\begin{figure*}
    \centering
    \includegraphics[width=16cm]{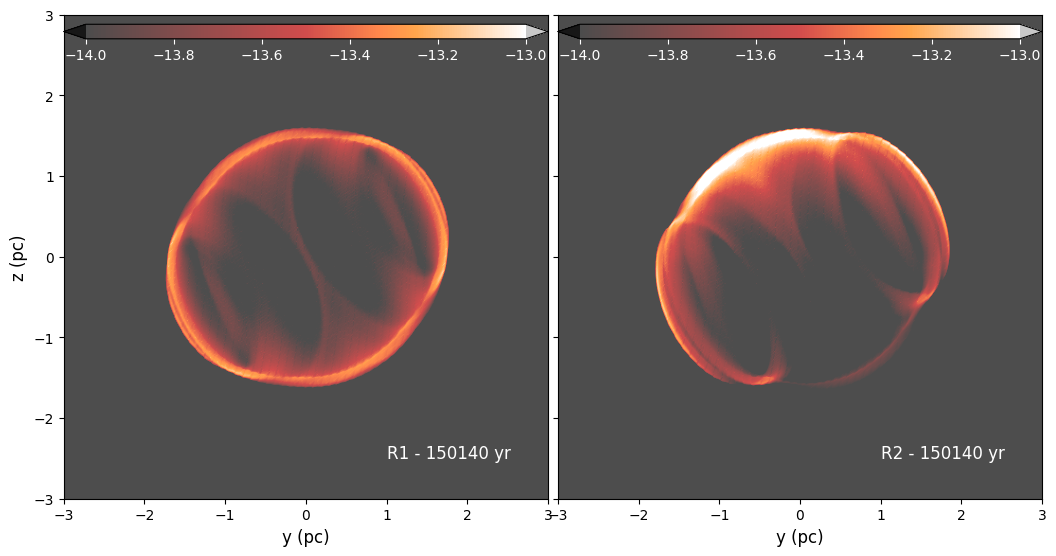}
    \caption{X-ray inverse Compton logarithm intensity maps in erg s$^{-1}$ cm$^{-2}$ sr$^{-1}$ for a 2-10 KeV band where the background radiation comes from $Q_L$. Model R1 on the left and model R2 on the right}
    \label{fig:InvCom_maps_smt}
\end{figure*}

Similarly but considering the radiative cooling of the MHD code as a photon source we build the IC maps illustrated in Figure~\ref{fig:InvCom_maps_smt}. There, the left hand side of the image, corresponding to the model R1, has a distribution that is similar to the H$\alpha$ but five orders of magnitude stronger, having peaks of intensity $j_{IC} \sim 10^{-13}$ erg s$^{-1}$ cm$^{-1}$ sr$^{-1}$ at the sides of the object. Here the difference between sides and caps is smaller. The top and bottom caps have an intensity $j_{IC} \sim 10^{-13.5}$ erg s$^{-1}$ cm$^{-1}$ sr$^{-1}$ but they show small regions with higher values. The right panel of the image shows the R2 model. The effects of an efficient cooling action are noticeable in the region interacting with the densest part of the background density gradient. The top cap of the object has a large region with intensity $j_{IC} > 10^{-13}$ erg s$^{-1}$ cm$^{-1}$ sr$^{-1}$ and it completely overshadows the intensity of the sides.

\subsubsection{Mixed emission}

\begin{figure*}
    \centering
    \includegraphics[width=16cm]{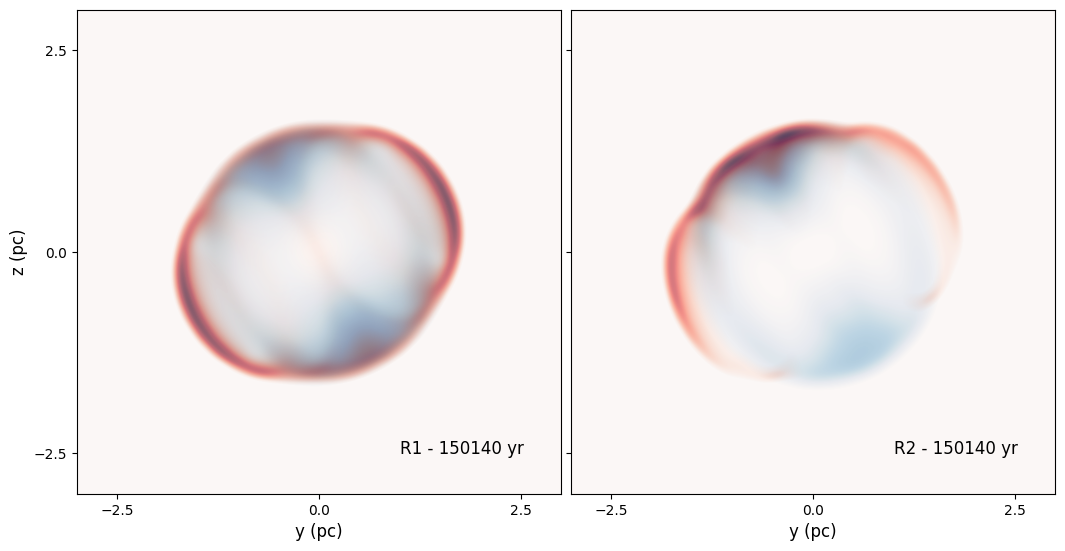}
    \caption{Overlapped maps of the weighted synchrotron (blue) and IC (red) emission due tu H$\alpha$ radiation for the R1 (left) and R2 (right) models.}
    \label{fig:Overlaped_maps}
\end{figure*}

A map of the multi-wavelength emission created is useful to further illustrate the mixed morphological nature of our object. In Figure \ref{fig:Overlaped_maps} we present a map of the overlapped emission obtained from our synthetic observations. Both emissivities, radio synchrotron in blue and IC X-ray in red, are normalized to their respective maxima. In these images the elliptical geometry of the object is evident. The asymmetrical synchrotron emission in both the R1 and R2 models is complemented by our modeled IC emission. The discrepancies between both emission methods are in good correspondence with their observational counterparts. 

\subsection{Comparison with observations}
Our simulations show that the anisotropy introduced by a stellar wind with a denser equatorial region is essential to reproduce the size and morphology observed in SNR G1.9+0.3. The critical feature of this astrophysical object is that the non-thermal emission in both radio and X-rays exhibit different brightness distributions. This characteristic asymmetrical barrel-shape morphology could be accentuated by adding jets to the pre-SN system, as was done in \cite{Akashi2018}.

We found that the magnetic field distribution is the most critical feature for generating synthetic synchrotron maps similar to the observed ones reported by \citet{reynolds2008} \citep[also][]{borkowski2017}. Due to the supernova remnant expanding into an elongated stellar wind bubble, there is low synchrotron emission from the regions called ears because the primary SNR shock wave is still inside of the stellar wind bubble, i.e. it is not directly interacting with the interstellar medium magnetic field (see Figure \ref{fig:Syncro_maps_smt}).

Several previous works report that the X-ray emission observed in SNR G1.9+0.3 is of non-thermal origin and assume that this emission is due to the synchrotron mechanism \citep{reynolds2008,borkowski2017,tsuji2021}. However, as \citet{borkowski2017} mentioned, it is intriguing that the brightness distribution in radio and X-rays do not coincide for this object, which does happen, for example, in the remnant of SN1006. \citet{borkowski2017} suggested that there must probably be another origin for the non-thermal X-ray emission observed in the G1.9+0.3 remnant. 

Therefore, we explored the role of the inverse Compton mechanism as a component of the non-thermal X-rays in this work. As mentioned before, this process requires two critical elements for it to take place: on the one hand, a relativistic electron distribution and, on the other, a radiation energy density $U_r$ that provides the photon pool that will scatter at X-ray frequencies. We employed the $\rm H\alpha$ emission and the radiative losses used by our code as proxies of this photon pool. Combining these elements, we generated IC maps which display a different brightness distribution than those obtained in the radio maps. Maps corresponding to the R1 (Fig. \ref{fig:InvCom_maps_smt}, left) run show higher emissions in the regions of the ears. In the case of run R2 (Fig. \ref{fig:InvCom_maps_smt}, right), we get a similar emission distribution but with an intensity growing in the direction of the imposed density gradient. This increase is even more remarkable when the energy density is given by radiative cooling since an increase in density produces an increase in radiative losses. The high IC emission obtained in the northeast for run R2 might occur because we impose a constant spectral index  (0.6) for each point belonging to the remnant. However, \citet{luken2020} showed that the NE of this astrophysical object has a higher spectral index value (larger than 0.8). To study the impact of this possibility, we carried out a test, performing an IC map with a spectral index of 0.8. In this case, the obtained synthetic maps showed IC emission in the NE remnant region lower than the previous one by one order of magnitude. Notwithstanding, the remarkable result is that we successfully reproduced the observed emission in both radio and X-ray frequencies. 

We successfully reproduce the X-ray ear features observed in this astrophysical object. Moreover, our synchrotron calculations explain the morphology and the distribution of magnetic field position angles. Aditionally, we include the radiative losses in our description, which are necessary when the shock wave collides with denser regions.

\section{Conclusions}\label{Sec:4}

In this study, we have modelled the structure of the type Ia supernova 
remnant G1.9+0.3, which displays an ears-like projected features and a mixed-morphology 
of its non-thermal radio and X-ray emission. We make use of three-dimensional 
magneto-hydrodynamical simulations complemented by computing synthetic maps 
for both synchrotron and inverse Compton emission. 

The adopted scenario suggests that the Type Ia SN occurred in the center of a peanut-shaped circumstellar bubble that possesses a density enhancement along the equatorial plane formed by intense mass outflows emanating from the progenitor system. Within the framework of Type Ia SNe, such dense and bipolar circumstellar structures are expected either by symbiotic binary progenitors in the single degenerated regime \citep[e.g][]{Chiotellis2012,Chiotellis2013,Broersen2014,Toledo-Roy2014} or by an episodic mass outflow during the planetary nebula phase of the white dwarf progenitor or by a common envelope episode of the parent stellar systems in double degenerate/core degenerate regime \citep[e.g][]{tsebrenko15b,2020Galax...8...38C,Soker2023}.

In this paper we worked with a scenario where the non-thermal X-ray emission from the ears of SNR G1.9+0.3 is produced by inverse Compton processes. This is different from the usually adopted framework, where all the non-thermal emission originates via the synchrotron process. The effects of two different IC proxies of the target photon fields are explored, each of them resulting in non-thermal X-ray emission maps with geometrically distinct distributions contrasting with the synchrotron radio maps. This methodology allowed us to build synthetic maps that reproduce the mixed morphology that characterizes this remnant.   

We found that the peculiar spatial distribution of the circumstellar medium is the key parameter for the blastwave to intercept an asymmetrically distributed material, produce the projected X-ray emission and to generate the observed mixed-morphology of G1.9+0.3. 
Additionally, the density gradient of the background interstellar medium accentuates the differences in synchrotron emission intensity between the NE and SW caps of the observations of G1.9+0.3. 

We observed an increased X-ray IC emission on the NE region mainly due to the increased H$\alpha$ (or radiative cooling) emission. In this particular region this can be lowered considering a higher spectral index (as \citealt{luken2020} showed by means of a spectral index distribution study), which would correspond to a shock that is being decelerated \citep[see][]{borkowski2017}. An in-situ estimation of the spectral index at each column could be useful for obtaining synthetic maps in the future.

Summarising, these results gave us an insight into two important considerations. First, to reproduce G1.9+0.3's ears it is essential to consider a remnant evolving inside an elongated stellar wind bubble. Second, the emission due to the IC process is able to reconcile the discrepancy between radio and X-ray observations. Different radiation sources should be taken into account to have a complete picture of the SNR G1.9+0.3 and to understand the appearance of mixed-morphology objects.

\section*{Acknowledgements}
Our colleague Alejandro Cristian Raga passed away on July 20, 2023, during the final stages of writing this paper. He was an outstanding scientist and a dear colleague and friend. We all miss him deeply. We thank the useful comments and suggestions given by the referee, which helped us to improve the previous version of this manuscript. MV is a doctoral fellow of CONICET, Argentina. MV and DOG acknowledge grant PICT 1046/2019 from ANPCyT to IAFE. PFV, ACR, AE, JCT-R, and KMV-R acknowledge financial support from PAPIIT-UNAM grants IG100422, IN113522, and IA103121. DOG and EMS are members of the Carrera del Investigador Cient\'\i fico of CO\-NI\-CET, Argentina. JCTR also acknowledges CONAHCYT grant CF-2023-I-2697. We thank Enrique Palacios (ICN-UNAM) for maintaining the Linux cluster, where the simulations were performed. The authors acknowledge the North-German Supercomputing Alliance (HLRN) for providing HPC resources that have contributed to the research results reported in this paper. 

\section*{Data Availability}
The data underlying this article will be shared on reasonable request to the corresponding author.



\bibliographystyle{mnras}
\bibliography{ref} 








\bsp	
\label{lastpage}
\end{document}